

\message
{JNL.TEX version 0.92 as of 6/9/87.  Report bugs and problems to Doug Eardley.}
\message
{This is X.G. Wen's copy}

\catcode`@=11
\expandafter\ifx\csname inp@t\endcsname\relax\let\inp@t=\input
\def\input#1 {\expandafter\ifx\csname #1IsLoaded\endcsname\relax
\inp@t#1%
\expandafter\def\csname #1IsLoaded\endcsname{(#1 was previously loaded)}
\else\message{\csname #1IsLoaded\endcsname}\fi}\fi
\catcode`@=12



\font\twelverm=cmr10 scaled 1200    \font\twelvei=cmmi10 scaled 1200
\font\twelvesy=cmsy10 scaled 1200   \font\twelveex=cmex10 scaled 1200
\font\twelvebf=cmbx10 scaled 1200   \font\twelvesl=cmsl10 scaled 1200
\font\twelvett=cmtt10 scaled 1200   \font\twelveit=cmti10 scaled 1200
\font\twelvesc=cmcsc10 scaled 1200  \font\twelvesf=amssmc10 scaled 1200
\skewchar\twelvei='177   \skewchar\twelvesy='60


\def\twelvepoint{\normalbaselineskip=12.4pt plus 0.1pt minus 0.1pt
  \abovedisplayskip 12.4pt plus 3pt minus 9pt
  \belowdisplayskip 12.4pt plus 3pt minus 9pt
  \abovedisplayshortskip 0pt plus 3pt
  \belowdisplayshortskip 7.2pt plus 3pt minus 4pt
  \smallskipamount=3.6pt plus1.2pt minus1.2pt
  \medskipamount=7.2pt plus2.4pt minus2.4pt
  \bigskipamount=14.4pt plus4.8pt minus4.8pt
  \def\rm{\fam0\twelverm}          \def\it{\fam\itfam\twelveit}%
  \def\sl{\fam\slfam\twelvesl}     \def\bf{\fam\bffam\twelvebf}%
  \def\mit{\fam 1}                 \def\cal{\fam 2}%
  \def\sc{\twelvesc}		   \def\tt{\twelvett}
  \def\sf{\twelvesf}
  \textfont0=\twelverm   \scriptfont0=\tenrm   \scriptscriptfont0=\sevenrm
  \textfont1=\twelvei    \scriptfont1=\teni    \scriptscriptfont1=\seveni
  \textfont2=\twelvesy   \scriptfont2=\tensy   \scriptscriptfont2=\sevensy
  \textfont3=\twelveex   \scriptfont3=\twelveex  \scriptscriptfont3=\twelveex
  \textfont\itfam=\twelveit
  \textfont\slfam=\twelvesl
  \textfont\bffam=\twelvebf \scriptfont\bffam=\tenbf
  \scriptscriptfont\bffam=\sevenbf
  \normalbaselines\rm}



\def\beginlinemode{\endmode
  \begingroup\parskip=0pt \obeylines\def\\{\par}\def\endmode{\par\endgroup}}
\def\beginparmode{\endmode
  \begingroup \def\endmode{\par\endgroup}}
\let\endmode=\par
{\obeylines\gdef\
{}}
\def\singlespace{\baselineskip=\normalbaselineskip}

\def\oneandahalfspace{\baselineskip=\normalbaselineskip
  \multiply\baselineskip by 3 \divide\baselineskip by 2}
\def\doublespace{\baselineskip=\normalbaselineskip \multiply\baselineskip by 2}

\newcount\firstpageno
\firstpageno=2
\footline={\ifnum\pageno<\firstpageno{\hfil}\else{\hfil\twelverm\folio\hfil}\fi}
\def\toppageno{\global\footline={\hfil}\global\headline
  ={\ifnum\pageno<\firstpageno{\hfil}\else{\hfil\twelverm\folio\hfil}\fi}}
\let\rawfootnote=\footnote		
\def\footnote#1#2{{\rm\singlespace\parindent=0pt\parskip=0pt
  \rawfootnote{#1}{#2\hfill\vrule height 0pt depth 6pt width 0pt}}}
\def\raggedcenter{\leftskip=4em plus 12em \rightskip=\leftskip
  \parindent=0pt \parfillskip=0pt \spaceskip=.3333em \xspaceskip=.5em
  \pretolerance=9999 \tolerance=9999
  \hyphenpenalty=9999 \exhyphenpenalty=9999 }
\def\dateline{\rightline{\ifcase\month\or
  January\or February\or March\or April\or May\or June\or
  July\or August\or September\or October\or November\or December\fi
  \space\number\year}}
\def\received{\vskip 3pt plus 0.2fill
 \centerline{\sl (Received\space\ifcase\month\or
  January\or February\or March\or April\or May\or June\or
  July\or August\or September\or October\or November\or December\fi
  \qquad, \number\year)}}


\hsize=6.5truein
\hoffset=0truein
\vsize=8.9truein
\voffset=0truein
\parskip=\medskipamount
\def\\{\cr}
\twelvepoint		
\doublespace		
\overfullrule=0pt	




\def\title			
  {\null\vskip 3pt plus 0.2fill
   \beginlinemode \doublespace \raggedcenter \bf}

\def\author			
  {\vskip 3pt plus 0.2fill \beginlinemode
   \singlespace \raggedcenter\sc}

\def\affil			
  {\vskip 3pt plus 0.1fill \beginlinemode
   \oneandahalfspace \raggedcenter \sl}

\def\abstract			
  {\vskip 3pt plus 0.3fill \beginparmode
   \oneandahalfspace ABSTRACT: }

\def\endtitlepage		
  {\endpage			
   \body}
\let\endtopmatter=\endtitlepage

\def\body			
  {\beginparmode}		

\def\head#1{			
  \goodbreak\vskip 0.5truein	
  {\immediate\write16{#1}
   \raggedcenter \uppercase{#1}\par}
   \nobreak\vskip 0.25truein\nobreak}

\def\beginitems{
\par\medskip\bgroup\def\i##1 {\item{##1}}\def\ii##1 {\itemitem{##1}}
\leftskip=36pt\parskip=0pt}
\def\enditems{\par\egroup}

\def\beneathrel#1\under#2{\mathrel{\mathop{#2}\limits_{#1}}}

\def\refto#1{$^{#1}$}		

\def\references			
  {\head{References}		
   \beginparmode
   \frenchspacing \parindent=0pt \leftskip=1truecm
   \parskip=8pt plus 3pt \everypar{\hangindent=\parindent}}

\def\referencesnohead   	
  {                     	
   \beginparmode
   \frenchspacing \parindent=0pt \leftskip=1truecm
   \parskip=8pt plus 3pt \everypar{\hangindent=\parindent}}

\gdef\refis#1{\item{#1.\ }}			

\gdef\journal#1, #2, #3, 1#4#5#6{		
    {\sl #1~}{\bf #2}, #3 (1#4#5#6)}		

\def\pr{\journal Phys. Rev., }

\def\prb{\journal Phys. Rev. B, }

\def\prl{\journal Phys. Rev. Lett., }

\def\np{\journal Nucl. Phys., }

\def\pl{\journal Phys. Lett., }

\def\endreferences{\body}

\def\figurecaptions		
  {\endpage
   \beginparmode
   \head{Figure Captions}
}

\def\endpage			
  {\vfill\eject}

\def\endpaper			
  {\endmode\vfill\supereject}


\def\heading				
  {\vskip 0.5truein plus 0.1truein	
   \beginparmode \def\\{\par} \parskip=0pt \singlespace \raggedcenter}

\def\subheading				
  {\vskip 0.25truein plus 0.1truein	
   \beginlinemode \singlespace \parskip=0pt \def\\{\par}\raggedcenter}

\def\tag#1$${\eqno(#1)$$}

\def\align#1$${\eqalign{#1}$$}

\def\aligntag#1$${\gdef\tag##1\\{&(##1)\cr}\eqalignno{#1\\}$$
  \gdef\tag##1$${\eqno(##1)$$}}

\def\endaligntag{}

\def\overset #1\to#2{{\mathop{#2}\limits^{#1}}}
\def\underset#1\to#2{{\let\next=#1\mathpalette\undersetpalette#2}}
\def\undersetpalette#1#2{\vtop{\baselineskip0pt
\ialign{$\mathsurround=0pt #1\hfil##\hfil$\crcr#2\crcr\next\crcr}}}


\def\ref#1{Ref.~#1}			
\def\Ref#1{Ref.~#1}			
\def\[#1]{[\cite{#1}]}
\def\cite#1{{#1}}
\def\(#1){(\call{#1})}
\def\call#1{{#1}}
\def\taghead#1{}
\def\frac#1#2{{#1 \over #2}}

\def\12{{1\over2}}

\def\ie{{\it i.e.,\ }}

\def\sla{\raise.15ex\hbox{$/$}\kern-.57em}
\def\leaderfill{\leaders\hbox to 1em{\hss.\hss}\hfill}
\def\twiddle{\lower.9ex\rlap{$\kern-.1em\scriptstyle\sim$}}
\def\bigtwiddle{\lower1.ex\rlap{$\sim$}}
\def\gtwid{\mathrel{\raise.3ex\hbox{$>$\kern-.75em\lower1ex\hbox{$\sim$}}}}
\def\ltwid{\mathrel{\raise.3ex\hbox{$<$\kern-.75em\lower1ex\hbox{$\sim$}}}}
\def\square{\kern1pt\vbox{\hrule height 1.2pt\hbox{\vrule width 1.2pt\hskip 3pt
   \vbox{\vskip 6pt}\hskip 3pt\vrule width 0.6pt}\hrule height 0.6pt}\kern1pt}
\def\tdot#1{\mathord{\mathop{#1}\limits^{\kern2pt\ldots}}}

\def\pmb#1{\setbox0=\hbox{#1}%
  \kern-.025em\copy0\kern-\wd0
  \kern  .05em\copy0\kern-\wd0
  \kern-.025em\raise.0433em\box0 }

\catcode`@=11
\newcount\r@fcount \r@fcount=0
\newcount\r@fcurr
\immediate\newwrite\reffile
\newif\ifr@ffile\r@ffilefalse
\def\w@rnwrite#1{\ifr@ffile\immediate\write\reffile{#1}\fi\message{#1}}

\def\writer@f#1>>{}
\def\referencefile{
  \r@ffiletrue\immediate\openout\reffile=\jobname.ref%
  \def\writer@f##1>>{\ifr@ffile\immediate\write\reffile%
    {\noexpand\refis{##1} = \csname r@fnum##1\endcsname = %
     \expandafter\expandafter\expandafter\strip@t\expandafter%
     \meaning\csname r@ftext\csname r@fnum##1\endcsname\endcsname}\fi}%
  \def\strip@t##1>>{}}

\def\citeall#1{\xdef#1##1{#1{\noexpand\cite{##1}}}}
\def\cite#1{\each@rg\citer@nge{#1}}	

\def\each@rg#1#2{{\let\thecsname=#1\expandafter\first@rg#2,\end,}}
\def\first@rg#1,{\thecsname{#1}\apply@rg}	
\def\apply@rg#1,{\ifx\end#1\let\next=\relax
\else,\thecsname{#1}\let\next=\apply@rg\fi\next}

\def\citer@nge#1{\citedor@nge#1-\end-}	
\def\citer@ngeat#1\end-{#1}
\def\citedor@nge#1-#2-{\ifx\end#2\r@featspace#1 
  \else\citel@@p{#1}{#2}\citer@ngeat\fi}	
\def\citel@@p#1#2{\ifnum#1>#2{\errmessage{Reference range #1-#2\space is bad.}%
    \errhelp{If you cite a series of references by the notation M-N, then M and
    N must be integers, and N must be greater than or equal to M.}}\else%
 {\count0=#1\count1=#2\advance\count1
by1\relax\expandafter\r@fcite\the\count0,%
  \loop\advance\count0 by1\relax
    \ifnum\count0<\count1,\expandafter\r@fcite\the\count0,%
  \repeat}\fi}

\def\r@featspace#1#2 {\r@fcite#1#2,}	
\def\r@fcite#1,{\ifuncit@d{#1}
    \newr@f{#1}%
    \expandafter\gdef\csname r@ftext\number\r@fcount\endcsname%
                     {\message{Reference #1 to be supplied.}%
                      \writer@f#1>>#1 to be supplied.\par}%
 \fi%
 \csname r@fnum#1\endcsname}
\def\ifuncit@d#1{\expandafter\ifx\csname r@fnum#1\endcsname\relax}%
\def\newr@f#1{\global\advance\r@fcount by1%
    \expandafter\xdef\csname r@fnum#1\endcsname{\number\r@fcount}}

\let\r@fis=\refis			
\def\refis#1#2#3\par{\ifuncit@d{#1}
   \newr@f{#1}%
   \w@rnwrite{Reference #1=\number\r@fcount\space is not cited up to now.}\fi%
  \expandafter\gdef\csname r@ftext\csname r@fnum#1\endcsname\endcsname%
  {\writer@f#1>>#2#3\par}}

\def\ignoreuncited{
   \def\refis##1##2##3\par{\ifuncit@d{##1}%
     \else\expandafter\gdef\csname r@ftext\csname
r@fnum##1\endcsname\endcsname%
     {\writer@f##1>>##2##3\par}\fi}}

\def\r@ferr{\endreferences\errmessage{I was expecting to see
\noexpand\endreferences before now;  I have inserted it here.}}
\let\r@ferences=\references
\def\references{\r@ferences\def\endmode{\r@ferr\par\endgroup}}

\let\endr@ferences=\endreferences
\def\endreferences{\r@fcurr=0
  {\loop\ifnum\r@fcurr<\r@fcount
    \advance\r@fcurr by 1\relax\expandafter\r@fis\expandafter{\number\r@fcurr}%
    \csname r@ftext\number\r@fcurr\endcsname%
  \repeat}\gdef\r@ferr{}\endr@ferences}


\let\r@fend=\endpaper\gdef\endpaper{\ifr@ffile
\immediate\write16{Cross References written on []\jobname.REF.}\fi\r@fend}

\catcode`@=12

\citeall\refto		
\citeall\ref		%
\citeall\Ref		%

\catcode`@=11
\newcount\tagnumber\tagnumber=0

\immediate\newwrite\eqnfile
\newif\if@qnfile\@qnfilefalse
\def\write@qn#1{}
\def\writenew@qn#1{}
\def\w@rnwrite#1{\write@qn{#1}\message{#1}}
\def\@rrwrite#1{\write@qn{#1}\errmessage{#1}}

\def\taghead#1{\gdef\t@ghead{#1}\global\tagnumber=0}
\def\t@ghead{}

\expandafter\def\csname @qnnum-3\endcsname
  {{\t@ghead\advance\tagnumber by -3\relax\number\tagnumber}}
\expandafter\def\csname @qnnum-2\endcsname
  {{\t@ghead\advance\tagnumber by -2\relax\number\tagnumber}}
\expandafter\def\csname @qnnum-1\endcsname
  {{\t@ghead\advance\tagnumber by -1\relax\number\tagnumber}}
\expandafter\def\csname @qnnum0\endcsname
  {\t@ghead\number\tagnumber}
\expandafter\def\csname @qnnum+1\endcsname
  {{\t@ghead\advance\tagnumber by 1\relax\number\tagnumber}}
\expandafter\def\csname @qnnum+2\endcsname
  {{\t@ghead\advance\tagnumber by 2\relax\number\tagnumber}}
\expandafter\def\csname @qnnum+3\endcsname
  {{\t@ghead\advance\tagnumber by 3\relax\number\tagnumber}}

\def\equationfile{%
  \@qnfiletrue\immediate\openout\eqnfile=\jobname.eqn%
  \def\write@qn##1{\if@qnfile\immediate\write\eqnfile{##1}\fi}
  \def\writenew@qn##1{\if@qnfile\immediate\write\eqnfile
    {\noexpand\tag{##1} = (\t@ghead\number\tagnumber)}\fi}
}

\def\callall#1{\xdef#1##1{#1{\noexpand\call{##1}}}}
\def\call#1{\each@rg\callr@nge{#1}}

\def\each@rg#1#2{{\let\thecsname=#1\expandafter\first@rg#2,\end,}}
\def\first@rg#1,{\thecsname{#1}\apply@rg}
\def\apply@rg#1,{\ifx\end#1\let\next=\relax%
\else,\thecsname{#1}\let\next=\apply@rg\fi\next}

\def\callr@nge#1{\calldor@nge#1-\end-}
\def\callr@ngeat#1\end-{#1}
\def\calldor@nge#1-#2-{\ifx\end#2\@qneatspace#1 %
  \else\calll@@p{#1}{#2}\callr@ngeat\fi}
\def\calll@@p#1#2{\ifnum#1>#2{\@rrwrite{Equation range #1-#2\space is bad.}
\errhelp{If you call a series of equations by the notation M-N, then M and
N must be integers, and N must be greater than or equal to M.}}\else%
 {\count0=#1\count1=#2\advance\count1
by1\relax\expandafter\@qncall\the\count0,%
  \loop\advance\count0 by1\relax%
    \ifnum\count0<\count1,\expandafter\@qncall\the\count0,%
  \repeat}\fi}

\def\@qneatspace#1#2 {\@qncall#1#2,}
\def\@qncall#1,{\ifunc@lled{#1}{\def\next{#1}\ifx\next\empty\else
  \w@rnwrite{Equation number \noexpand\(>>#1<<) has not been defined yet.}
  >>#1<<\fi}\else\csname @qnnum#1\endcsname\fi}

\let\eqnono=\eqno
\def\eqno(#1){\tag#1}
\def\tag#1$${\eqnono(\displayt@g#1 )$$}

\def\aligntag#1\endaligntag
  $${\gdef\tag##1\\{&(##1 )\cr}\eqalignno{#1\\}$$
  \gdef\tag##1$${\eqnono(\displayt@g##1 )$$}}

\def\eqalignno#1{\displ@y \tabskip\centering
  \halign to\displaywidth{\hfil$\displaystyle{##}$\tabskip\z@skip
    &$\displaystyle{{}##}$\hfil\tabskip\centering
    &\llap{$\displayt@gpar##$}\tabskip\z@skip\crcr
    #1\crcr}}

\def\displayt@gpar(#1){(\displayt@g#1 )}

\def\displayt@g#1 {\rm\ifunc@lled{#1}\global\advance\tagnumber by1
        {\def\next{#1}\ifx\next\empty\else\expandafter
        \xdef\csname @qnnum#1\endcsname{\t@ghead\number\tagnumber}\fi}%
  \writenew@qn{#1}\t@ghead\number\tagnumber\else
        {\edef\next{\t@ghead\number\tagnumber}%
        \expandafter\ifx\csname @qnnum#1\endcsname\next\else
        \w@rnwrite{Equation \noexpand\tag{#1} is a duplicate number.}\fi}%
  \csname @qnnum#1\endcsname\fi}

\def\ifunc@lled#1{\expandafter\ifx\csname @qnnum#1\endcsname\relax}

\let\@qnend=\end\gdef\end{\if@qnfile
\immediate\write16{Equation numbers written on []\jobname.EQN.}\fi\@qnend}

\catcode`@=12


\def\ie{{\it i.e.,\ }}

\def\>{\rangle}
\def\<{\langle}
\def\o{\over}

\def\t{\tilde}

\def\slD{\raise.15ex\hbox{$/$}\kern-.57em\hbox{$D$}}
\def\dsl{\raise.15ex\hbox{$/$}\kern-.57em\hbox{$\Delta$}}
\def\slp{{\raise.15ex\hbox{$/$}\kern-.57em\hbox{$\partial$}}}
\def\nsl{\raise.15ex\hbox{$/$}\kern-.57em\hbox{$\nabla$}}
\def\sla{\raise.15ex\hbox{$/$}\kern-.57em\hbox{$\rightarrow$}}
\def\slla{\raise.15ex\hbox{$/$}\kern-.57em\hbox{$\lambda$}}
\def\slb{\raise.15ex\hbox{$/$}\kern-.57em\hbox{$b$}}
\def\lnp{\raise.15ex\hbox{$/$}\kern-.57em\hbox{$p$}}
\def\lnk{\raise.15ex\hbox{$/$}\kern-.57em\hbox{$k$}}
\def\lnK{\raise.15ex\hbox{$/$}\kern-.57em\hbox{$K$}}
\def\lnq{\raise.15ex\hbox{$/$}\kern-.57em\hbox{$q$}}

\def\de{{\delta}}

\def\si{{\sigma}}

\def\cA{{\cal A}}

\def\cS{{\cal S}}

\def\part{\partial}

\def\abs{
         \vskip 3pt plus 0.3fill\beginparmode
         \doublespace ABSTRACT:\ }


\title
Chiral Operator Product Algebra
Hidden in Certain Fractional Quantum Hall Wave Functions

\author Xiao-Gang Wen

\affil
Department of Physics
MIT
77 Massachusetts Avenue
Cambridge, MA 02139

\author Yong-Shi Wu

\affil
Department of Physics
University of Utah
Salt Lake City, UT 84112

\abs{
In this paper we study the conditions under which
an N-electron wave function for a fractional
quantum Hall (FQH) state can be viewed as N-point
correlation function in a conformal field theory (CFT).
Several concrete examples are presented to illustrate,
when these condition are satisfied, how to ``derive''
or ``uncover'' relevant operator algebra in
the associated CFT from the FQH wave function.
Besides the known Pfaffian state, the states
studied here include three d-wave paired states,
one for spinless electrons and two for spin-1/2
electrons (one of them is the Haldane-Rezayi state).
It is suggested that the non-abelian topological order
hidden in these states can be characterized by their
associated chiral operator product algebra,
from which one may infer the quantum
numbers of quasi-particles and calculate their wave functions.
}

\endtopmatter

\head{1. Introduction}

Fractional quantum Hall (FQH) states contain extremely
rich internal structures --as first revealed by the existence of
the hierarchical states\refto{h}-- which one may call topological
orders.\refto{rd} While the topological orders in the abelian
FQH states are known to be characterized
by symmetric integer matrices,\refto{mat} the characterization
of the non-abelian FQH states, i.e. those supporting
quasiparticle excitations obeying non-abelian braid statistics,
remains an open question. This paper is devoted to the study
of the possibility of characterizing some {\it non-abelian}
FQH states by conformal field theory (CFT).

In the literature there have been evidences suggesting that at least
some FQH states are closely related to conformal field theories.
First, on one hand, the effective theory for abelian FQH
states\refto{eff,rd,mat} and a class of non-abelian FQH states generated
by the Kac-Moody algebra\refto{W} are topological Chern-Simons (CS)
theories, which are known to be connected to CFT.\refto{witten} On the
other hand, Moore and Read\refto{MR}, and also Fubini\refto{Fubini}
independently, have shown that some FQH wave functions can be formally
expressed as correlations of primary fields (or vertex operators) in
certain CFT.  In particular, the correspondence between usual Laughlin
wave functions and correlations in the $c=1$ U(1) CFT's has been
recently verified\refto{Naples} even for the genus-one case when the
FQH state is put on a torus.\refto{HR1} Furthermore, it has been
shown that the quantum numbers of quasiparticles supported by some
FQH states are determined by appropriate vertex or disorder
operators.\refto{MR,W} Also the ground state degeneracy and the
associated non-abelian Berry's phases of FQH states on torus, as shown
in \ref{rd,MR,Naples,cg}, are closely related to conformal blocks and
the associated modular transformations in CFT. Especially, the gapless
edge excitations of FQH states, by themselves, form chiral
CFT.\refto{edge,W,Wp} All these suggest that at least for some FQH
states, the topological order hidden in their wave functions
may be characterized by CFT, and the knowledge of the latter
is useful in studying topological properties of the FQH state,
such as the quantum numbers of quasiparticles, ground state degeneracy
and spectrum of edge states.  This approach should be helpful
particularly for theoretically exploring non-abelian FQH states which
are not well understood yet.

It is well-known that when restricted to the first
Landau level, apart from an exponential Gaussian factor,
a FQH state is described by a holomorphic function
$\Phi (z_i)$, where $z_i=x_i+iy_i$ is the complex
coordinate of the $i$-th electron. So the idea
has been to use correlation functions in a (chiral) CFT,
which are always holomorphic, to generate new non-abelian
FQH wave functions in the first Landau level.
However, in this approach the starting CFT is
chosen somewhat {\it a priori}; several non-abelian
FQH states constructed in this way\refto{MR,W}
awaits for their experimental realization. On
the other hand, there are FQH states whose many-body
wave function are constructed from physical
considerations, such as the Haldane-Rezayi
wave function for the $\nu=5/2$ state.
Given a FQH state as such, a more natural
approach seems to be first to ask whether
we can identify the wave function with
correlations in a CFT and then, if the answer is
yes, to use the operator (product) algebra
of the CFT to characterize the topological order
hidden in the wave function.

In this paper we will study some conditions under which a
given FQH wave function can be identified as correlation
functions in a CFT. Then we demonstrate how to derive
(the relevant operator algebra in) the associated CFT
from the FQH wave function, using the well-understood
Pfaffian state as an illustrative example.
Using this method, we further construct
the operator algebra in the CFT associated to
three d-wave paired FQH states respectively,
one for spinless electrons and the other two for spin-1/2
electrons (one of them is the Haldane-Rezayi state).
The CFT description allows us to study the structure of
quasiparticle excitations in the three d-wave FQH states.
The study of other physical consequences, such as the
ground state degeneracy on torus and the spectrum of
edge states, is in progress and is
left to future publication.

We would like to point out that once a FQH wave function
is written as correlation function in a certain CFT,
then the wave function can also be expressed as a
correlation in infinity many other CFT which contain
the original CFT. Thus one needs to introduce a concept of
minimal CFT in order to define a precise relation between
the CFT and FQH wave function. The minimal CFT for a
FQH state not only reproduces the wave function,
it is also contained in any other CFT that reproduces the
wave function. Some times it is not hard to guess a CFT
that reproduces a certain wave function. It is often difficult
to tell whether the guessed CFT is the minimal one or not.
The CFT constructed in this
paper is automatically the minimal one. This is because,
by construction, the CFT that we obtained contains the
minimal set
of primary fields that are needed to construct the
wave function. Furthermore the energy-momentum
tensor of the minimal CFT can be expressed in terms of the
operator that generate the wave function. As we will
see in a future publication that the minimal
CFT are closely related to the edge excitations
of the corresponding FQH state.

\head{2. General discussion and the Pfaffian state}

In general, the holomorphic part of a FQH many-body
wave function consists of two factors. One of them
is of usual Laughlin-Jastrow form,
$\prod_{i<j} (z_i-z_j)^r$ ($r$ is a fraction or
integer), which we call the $U(1)$ part. The connection of
this part to the $U(1)$ CFT is well-understood.
Here we concentrate on the other part, which is
not of the Laughlin-Jastrow form. We call it
the non-abelian part, because it is this part that
is presumably responsible to the
appearance of non-abelian statistics for quasiparticles.

We would like to address the following questions:
A) Given a holomorphic $N$-body FQH wave
function, is there a CFT whose $N$-point correlation
(for {\it arbitrary} $N$) reproduces the FQH wave function.
B) If such a CFT exists,
how to derive it (or its conformal operator algebra)
from the FQH wave function.
(For readers who are not familiar with CFT, we recommend
the \ref{book} and the classical paper of Belavin,
Polyakov and Zamolochikov.\refto{BPZ})

Mathematically, the affirmative
answer to question A) means the following.
Viewing the FQH wave function $\Phi$
as a correlation of a certain field
operator $\psi$ in a 2D quantum field theory,
$$
\Phi(z_i)=\< \prod \psi(z_i) \>, \ \ \ z_i=x_i+iy_i
\eqno(1)$$
we can find an operator $T(z)$ independent of $\bar z$
(called the energy-momentum tensor) such that
the FQH wave function satisfies the
conformal Ward identity\refto{BPZ}
$$
\< T(z) \prod_i \psi(z_i) \>=\sum_k \left( {h\o (z -z_k)^2}
+{1\o z -z_k} \part_{z_k} \right) \< \prod_i \psi(z_i) \>
\eqno(2)$$
with a real constant $h$, called the conformal
dimension of the (primary) field $\psi$.
The energy-momentum tensor should also satisfy
the Virasoro algebra, which is given, in the form of operator
product expansion (OPE), by
$$T(z_1)T(z_2)={c/2\o (z_1-z_2)^4}+ {2\o (z_1-z_2)^2} T(z_2)
+{1\o z_1-z_2}\part_{z_2} T(z_2) +O(1)
\eqno(3)$$
The real constant $c$ in \(3) is called the central charge.
In terms of correlation function, \(3) can be
written as\refto{BPZ}
$$\eqalign{
&\< T(z)T(\t z) \prod_i \psi(z_i) \> \cr
=&\sum_k \left( {h\o (z -z_k)^2}
 +{1\o z -z_k} \part_{z_k} \right)
\< T(\t z) \prod_i \psi(z_i) \>\cr
&+\left( {2\o (z -\t z)^2}+{1\o z -\t z} \part_{\t z} \right)
 \< T(\t z) \prod_i \psi(z_i) \>
 +{c\o 2(z-\t z)^4} \< \prod_i \psi(z_i) \> \cr}
\eqno(4)$$

In general the energy-momentum tensor $T(z)$
may or may not be contained in the OPE of the
field $\psi$. If the former is true, this
additional information will allow us
to express the conditions \(2) and \(4) as
conditions on the wave function $\Phi$. Even in this case,
a general solution to the question A) should be
quite complicated. So in this paper we restrict
ourselves to a simpler question: A')
Given a holomorphic FQH wave function, can we write it as
the correlations of a primary field $\psi$ in a CFT,
which has the following OPE
$$
\psi(z_1)\psi(z_2)={1\o (z_1-z_2)^{2h}}
\left(1+{2h\o c}(z_1-z_2)^2 T(z_2) + O( (z_1-z_2)^3) \right)
\eqno(5)$$
Here the additional simplifying assumption is that the OPE of $\psi$
does not mix with dimension 1 and 2 primary fields.

If the answer to A') is yes, \(5) implies that
the two-body wave function should have the
form $\Phi=1/(z_1-z_2)^{2h}$. Thus the
constant $h$ (the dimension of $\psi$) can be
easily read off from the two-body wave function.
\(5) also implies that
$$T(z)={c\o 2h} \lim_{z_1\to z}
\left(\psi(z_1)\psi(z)- {1\o (z_1-z)^{2h}} \right) (z_1-z)^{2h-2}
\eqno(6)$$
Substituting \(6) in to \(2) and \(4), we find that
\(2) and \(4) become conditions on the wave function $\Phi$,
which relate the $N$-body wave function to the $(N-2)$- and
$(N-4)$-body wave functions. Thus the question A' can be answered
by explicitly checking whether the wave function $\Phi$
satisfies these conditions or not.

Let us now apply the above general discussion to a simple example,
namely the p-wave paired FQH state for spinless
electrons discussed in \ref{MR}. The total wave
function is given by
the product of a Pfaffian wave function
$\Phi_{Pf}$ and a Laughlin wave function $\Phi_m$
$$\eqalign{
\Phi_p=&\Phi_{Pf}\Phi_m \cr
\Phi_{Pf}=&\cA({1\o (z_1-z_2)}{1\o (z_3-z_4)}...) \cr
\Phi_m=& \left( \prod_{i<j} (z_i-z_j)^m \right)
 e^{-{1\o 4} \sum_i |z_i|^2} \cr}
\eqno(7)$$
where $\cA$ is the antisymmetrization operator;
$m$ is an even integer. The Pfaffian wave function
is the exact incompressible ground state of a
three-body Hamiltonian.\refto{GWW,Wp,RR}

It is known that the Laughlin part of the wave function
can be written as correlation
of a vertex operator $e^{i\sqrt{m}\phi}$ in the
Gaussian model.\refto{MR,Fubini} (Here we adopt the normalization
$\< e^{i\phi(z)} e^{i\phi(0)}\>=1/z$.)
We would like to ask
whether the Pfaffian part can be written as a correlation of
certain operator $\psi$ in an appropriate CFT.

{}From the two point function
$\Phi_{Pf}=1/(z_1-z_2)$, we find that
$\psi$ has a dimension $h=1/2$.
$\Phi_{Pf}$ also has the following property: As $z_1\to z_2$
$$\Phi_{Pf}(z_1,..)={\Phi_{Pf}(z_3,..)\o z_1-z_2}+O(z_1-z_2)
$$
Note that no terms are of one power of $z_1-z_2$ higher than the
leading singular term. This implies that \(6)
defines a valid operator that has well defined correlations with
$\psi$'s.
Thus one may try to introduce the
energy-momentum tensor through \(6), which leads to
the identification
$$
 \< T(z_2) \prod_{i=3} \psi(z_i) \>
\equiv c \lim_{z_1\to z_2}\left( \Phi_{Pf}(z_1,z_2,...)
-{1\o z_1-z_2}   \Phi_{Pf}(z_3,z_4,...) \right) {1\o z_1-z_2}.
\eqno(a)$$
Thus \(2) and \(4) are reduced to the following
conditions on the Pfaffian wave function:
$$\eqalign{
& c \lim_{z_1\to z_2}\left( \Phi_{Pf}(z_1,z_2,...)
-{1\o z_1-z_2}   \Phi_{Pf}(z_3,z_4,...) \right) {1\o z_1-z_2} \cr
=& \sum_{k=3} \left( {1/2 \o (z_2 -z_k)^2}
   +{1\o z_2 -z_k} \part_{z_k} \right) \Phi_{Pf}(z_3,z_4,...) \cr}
\eqno(8)$$
and
$$\eqalign{
&c^2 \lim_{z_1\to z_2, z_3\to z_4}
\Big(
\Phi_{Pf}(z_1,z_2,...)-{\Phi_{Pf}(z_3,z_4,...)\o z_1-z_2}
-{\Phi_{Pf}(z_1,z_2,z_5...)\o z_3-z_4} \cr
&\ \ \ \ \ \ \ \ \ \ \ \ \ \
+{\Phi_{Pf}(z_5,z_6...)\o (z_1-z_2)(z_3-z_4)}
\Big){1\o (z_1-z_2)(z_3-z_4)} \cr
=&\sum_{k=5} \left( {1/2\o (z_2 -z_k)^2}
 +{1\o z_2 -z_k} \part_{z_k} \right)
 \< T(z_4) \prod_{i=5} \psi(z_i) \>\cr
&+\left( {2\o (z_2 -z_4)^2}+{1\o z_2-z_4} \part_{z_4} \right)
 \< T(z_4) \prod_{i=5} \psi(z_i) \>
 +{c/2\o (z_2-z_4)^4} \Phi_{Pf}(z_5,z_6...) \cr}
\eqno(9)$$
where $\< T(z_4) \prod_{i=5} \psi(z_i) \>$ is given by \(a).
It is straightforward to verify that the limits
in \(8) and \(9) are indeed finite, and both \(8) and \(9)
are satisfied by the Pfaffian wave function \(7)
if the central charge $c$ is chosen to be $1/2$. Thus the
validity of \(8) and \(9) assures us that
$\Phi_{Pf}$ can be identified with correlation of a
dimension-$1/2$ field in a $c=1/2$ CFT. Indeed,
the Pfaffian wave function was first constructed\refto{MR}
as correlations in the Ising model which has $c=1/2$.

To give a complete description of a CFT,
we also need to know the operator (product) algebra, or OPE,
for all primary fields\refto{BPZ}. In general the OPE
of the $\psi$ operator may generate new operators.
Thus $\psi$ can be viewed as a generator
which generates a closed operator algebra,
which will be called the center algebra.
It is plausible that the properties of this
operator algebra (such as structure constants,
correlation of new operators generated by $\psi$, etc.)
are determined by the correlations of $\psi$'s.
Then one can use the wave functions to calculate
the structure constants of the center algebra.
We note that the center algebra is always a semilocal
algebra. This is directly related to the single-valueness
of the electron wave function.

For the Pfaffian wave function, since we already
know that the wave function coincides with the correlation
of the dimension-$1/2$ field in the
Ising model, the knowledge of the latter\refto{BPZ}
can be used to determine the operator algebra
generated by $\psi$, which is simply
$$
\psi(z_1)\psi(z_2) \sim {1\o z_1-z_2}
\eqno(10)$$
\ie the OPE of $\psi$ does not generate
any new operators. Thus the center algebra for the
Pfaffian wave function is given by
$$ 1\times 1    \sim 1,\ \ \
   1\times \psi \sim \psi,\ \ \
\psi\times \psi \sim 1.
\eqno(10a)$$

We would like to point out that the structure of the operator
algebra generated by $\psi$ is very important.
This structure reflects the structure of internal
correlation of the wave function. Instead
of directly working with the wave function, we can study the
internal structure of FQH state by studying the associated
center algebra. The topological order of a
quantum Hall wave function can be determined by the center
 algebra. This is the heart
of the operator (or CFT) approach to the FQH wave function.

The CFT description is very useful in the study of the
properties of quasiparticles. To construct the quasiparticles
in the p-wave paired
FQH state, we need to extend the center algebra
generated by $\psi$ to include the disorder operator
$\si$, which is known to have dimension $1/16$ in
the Ising model. The extended operator algebra is generated by
$\psi$ and $\si$ which has the following fusion rules
$$\eqalign{
&
\psi\times \psi \sim 1,\ \ \
 \si\times \si  \sim 1,\ \ \
 \nu\times \nu  \sim 1,\ \ \ \cr
&
\psi\times \si  \sim \nu,\ \ \
\psi\times \nu  \sim \si,\ \ \
 \si\times \nu  \sim \psi. \cr}
\eqno(10b)$$
The quasiparticle excitations
can be expressed in terms of the known Ising
correlations of the $\si$ operators,
which in turn enable us to determine quantum numbers
of the quasiparticles. Detailed discussion can be
found in \ref{MR}. Note that any new disorder operator
in the extended algebra must be semilocal
with respect to $\psi$,
so that one can use them (together with appropriate
U(1) part) to construct single-valued electron
wave functions.\refto{MR,W}

\head{3. The D-wave Paired State for Spinless Electrons}

The next-to-simplest example is the d-wave paired FQH state
for spinless electrons, which is a natural generalization of
the Pfaffian or p-wave paired state. The total wave function for
this state is of a form similar to that of \(7):
$$\eqalign{
\Phi=&\Phi_{d}\Phi_m \cr
\Phi_{d}=&\cS({1\o (z_1-z_2)^2}{1\o (z_3-z_4)^2}...) \cr
\Phi_m=& \left( \prod_{i<j} (z_i-z_j)^m \right)
 e^{-{1\o 4} \sum_i |z_i|^2} \cr}
\eqno(7a)$$
The differences with \(7) are: first, the negative
power of paired $z_i-z_j$ in $\Phi_{d}$ is two rather than
one and, secondly, the symmetrization $\cS$ replaces the
antisymmetrization $\cA$. Thus $m$ must be an odd integer,
to make the total wave function $\Phi$ anti-symmetric.

Let us first analyze the structure of zeros of this wave
function assuming, for simplicity, $m=3$. Let
$z_1=z_3+\de_1$ and $z_2=z_3+\de_2$, we find $\Phi$ has the
following expansion
$$\Phi=\sum_{k=odd}(\de_2)^k\sum_l (\de_1)^l A_{kl}(z_3,z_4,...)
\eqno(7b)$$
One can directly check that the coefficients
$$A_{12}=A_{14}=A_{33}=0
\eqno(7c)$$
Therefore $\Phi$ is the exact ground state of Hamiltonian $H$
with the following three-body potential
$$\eqalign{
V=&V_1 \part_{z_2^*}    \de(z_2-z_3) \part_{z_2}
	\part_{z_1^*}^2  \de(z_1-z_3) \part_{z_1}^2\cr
   &+V_2 \part_{z_2^*}    \de(z_2-z_3) \part_{z_2}
	\part_{z_1^*}^4  \de(z_1-z_3) \part_{z_1}^4
   +V_3 \part_{z_2^*}^3  \de(z_2-z_3) \part_{z_2}^3
	\part_{z_1^*}^3  \de(z_1-z_3) \part_{z_1}^3\cr}
\eqno(7d)$$
This is because the Hamiltonian $H$ is positive definite
if $V_i>0$, and $\Phi$ is a zero-energy state of $H$.\refto{commH}
We have checked
numerically that $\Phi$ is indeed the non-degenerate
ground state of $H$ (on a sphere) with a finite energy gap.
This implies that the constraint \(7c)
uniquely fixes the wave function.

Now the two-point function is $\Phi_{d}=1/(z_1-z_2)^2$. Thus
$\psi$ is of dimension $h=1$. Also $\Phi_{d}$ has
the property that as $z_1\to z_2$,
$$\Phi_{d}(z_1,..)={\Phi_{d}(z_3,..)\o (z_1-z_2)^2}+O(1)
$$
Again there are no terms of one power of $z_1-z_2$
higher than the leading singular term. So \(6) makes
perfect sense and can be used to define the energy-momentum
tensor. This leads to the identification
$$
 \< T(z_2) \prod_{i=3} \psi(z_i) \>
\equiv {c\o 2} \lim_{z_1\to z_2}\left( \Phi_{d}(z_1,z_2,...)
-{1\o (z_1-z_2)^2}   \Phi_{d}(z_3,z_4,...) \right).
\eqno(b)$$
Thus the conformal Ward identities \(2) and \(4)
are reduced to the following
conditions on the wave function $\Phi_{d}$:
$$\eqalign{
& {c/2} \lim_{z_1\to z_2}\left( \Phi_{d}(z_1,z_2,...)
-{1\o (z_1-z_2)^2} \Phi_{d}(z_3,z_4,...) \right) \cr 
=& \sum_{k=3} \left( {1\o (z_2 -z_k)^2}
   +{1\o z_2 -z_k} \part_{z_k} \right) \Phi_{d}(z_3,z_4,...) \cr}
\eqno(b1)$$
and
$$\eqalign{
&{c^2\o 4} \lim_{z_1\to z_2, z_3\to z_4}
\Big(
\Phi_{d}(z_1,z_2,...)-{\Phi_{d}(z_3,z_4,...)\o (z_1-z_2)^2}
-{\Phi_{d}(z_1,z_2,z_5...)\o (z_3-z_4)^2} \cr
&\ \ \ \ \ \ \ \ \ \ \ \ \ \
+{\Phi_{d}(z_5,z_6...)\o (z_1-z_2)^2 (z_3-z_4)^2}\Big)\cr
=&\sum_{k=5} \left( {1\o (z_2 -z_k)^2}
 +{1\o z_2 -z_k} \part_{z_k} \right)
 \< T(z_4) \prod_{i=5} \psi(z_i) \>\cr
&+\left( {2\o (z_2 -z_4)^2}+{1\o z_2-z_4} \part_{z_4} \right)
 \< T(z_4) \prod_{i=5} \psi(z_i) \>
 +{c/2\o (z_2-z_4)^4} \Phi_{d}(z_5,z_6...) \cr}
\eqno(b2)$$
where $\< T(z_4) \prod_{i=5} \psi(z_i) \>$ is given by \(b).
One can directly check that both \(b1) and \(b2) are indeed
satisfied by the wave function $\Phi_{d}$ if we choose the
central charge $c=1$.

This implies that $\Phi_{d}$ can be viewed as a
correlation of the dimension-one ($h=1$) primary field
$\psi$ in a $c=1$ CFT. Note that these values of
the pair $(h,c)$ satisfies the Kac formula
with $n=3, m=1$:
$$ h_{(n,m)}= h_{0} + {1\o 4}
(\alpha_{+} n +\alpha_{-} m)^2,
\eqno(Kac)$$
where
$$ h_{0}= {1\o 24} (c-1),\;\; \alpha_{\pm}=
{\sqrt{1-c}\pm \sqrt{25-c} \o \sqrt{24}}.
\eqno(b3)$$
Thus $\psi$ is a degenerate primary field , whose
correlation (or the wave function $\Phi_{d}$)
should satisfy a third-order differential
equation\refto{BPZ}:
$$\eqalign{
\Big\{
& {1\o 2} {\partial^3 \o \partial z^3}
-\sum_{i=1} {2\o (z-z_i)^3} -\sum_{i=1}
{1\o (z-z_i)^2} {\partial\o \partial z_i}\cr
& -\sum_{i=1} {2\o (z-z_i)^2} {\partial\o \partial z}
-\sum_{i=1} {2\o (z-z_i)}
{\partial^2\o \partial z \partial z_i}
\Big\} \Phi_d (z, z_1,z_2,\cdots)=0. \cr}
\eqno(b4)$$
Indeed, one can explicitly verify that this
equation is satisfied by $\Phi_{d}$ given by
\(b). This further confirms that $\Phi_d$ can be
written as correlation in a $c=1$ CFT. In fact
we have directly checked that $\Phi_d$ is the correlation
of the $U(1)$ current in the Gaussian model.

Introducing the electron operator
$$\psi_e(z)=\psi(z) e^{i\sqrt{m} \phi(z)}
\eqno(b5)$$
we find that the wave function $\Phi$ in \(7a) can be written as
$$\Phi=\<\prod_i \psi_e(z_i) e^{-i{1\o \sqrt{m} }\int d^2 z \phi} \>
\eqno(b6)$$
Now the constraint \(7c) on zeros of $\Phi$
becomes a consequence of the OPE of
$\psi_e$. Let $z_1=z_3+\de_1$ and $z_2=z_3+\de_2$ and assume $m=3$,
we find
$$\eqalign{
   & \psi_e(z_1) \psi_e(z_2) \psi_e(z_3) \cr
=  & \psi_e(z_1) \left( \de_2 e^{2i\sqrt{m} \phi(z_3)}
      +(\de_2)^3 2T(z_3) e^{2i\sqrt{m} \phi(z_3)}\right)+...\cr
=  & \de_2 (\de_1)^6 \psi(z_3) e^{3i\sqrt{m} \phi(z_3)}
      +(\de_2)^3 (\de_1)^4 2\psi(z_3) e^{3i\sqrt{m} \phi(z_3)}+...\cr}
\eqno(b7)$$
The lower-order zeros are absent as described by \(7c).

Now let us discuss quasiparticle excitations in the d-wave
paired state. First we notice that the following operator
algebra
$$\eqalign{
\part_{\bar z}\psi=& 0\cr
[ \psi(z_1),\psi(z_2)]=& 0 \cr
\psi(z)\psi(0)=& {1\o z^2}+O(1) \cr}
\eqno(b20)$$
completely determines the correlations between $\psi$. This is
because the algebra \(b20) completely determines the poles and their
residues in the correlation. Being a holomorphic function,
the correlation is thus uniquely determined and
turns out to be none other than $\Phi_{d}$.

To study the quasiparticles in this FQH state,
we need to extend the
operator algebra \(b20) to include the
disorder operator $\eta$.
We require the correlation between $\psi$ and $\eta$ to acquire
a minus sign as we move $\psi$ around $\eta$. Other phases
are not allowed, because $\psi^2\sim 1$ and moving
a pair of $\psi$ around $\eta$ should not give any phase.
Thus we may try the following OPE between $\psi$ and $\eta$:
$$\psi(z)\eta(0)\sim z^{-{1\o 2}} (\t\eta(0)+O(z))
\eqno(b21)$$
where $\t\eta$ is some other operator.

To calculate the correlations between $\psi$'s and
two $\eta$ operators\refto{com}
 we note that, as a consequence of \(b20) and \(b21),
$$\prod_{m=1,2} (z_1-u_m)^{1/2}
\<\prod_{i=1}\psi(z_i)\prod_{m=1,2} \eta(u_m)\>
\eqno(b22)$$
is a holomorphic function of $z_1$ which has only poles at
$z_1=z_i$, $i=2,3,..$.
We also note that, as a function of $z_1$,
$$\Phi_q(z_1..;u_1,u_2)\equiv
\<\prod_{i=1}\psi(z_i) \prod_{m=1,2} \eta(u_m)\>
\eqno(b23)$$
has a pure second-order pole (no first-order pole)
at $z_1=z_i$ with a residue
$\Phi_q(\hat z_1,\hat z_i)$, where $\Phi_q(\hat z_1,\hat z_i)$
is $\Phi_q(z_1..;u_1,u_2)$ with variables $z_1$ and $z_i$ removed.
This implies that the second-order pole at $z_1=z_i$ in \(b22)
has a residue $\Phi_q(\hat z_1,\hat z_i)\prod_{m=1,2} (z_i-u_m)^{1/2}$
and the first-order pole has a residue
$\part_{z_1} \Phi_q(\hat z_1,\hat z_i)\prod_{m=1,2} (z_1-u_m)^{1/2}$
evaluated at $z_1=z_i$.
We obtain the following relation for $\Phi_q$:
$$\eqalign{
 &\Phi_q(z_1..;u_1,u_2) \cr
=&\prod_{m=1,2} (z_1-u_m)^{-1/2}\sum_{k=2}
\left({1\o (z_1-z_k)^2}+{1\o 2(z_1-z_k)}\sum_{m=1,2} {1\o z_k-u_m}\right) \cr
 &\times \prod_{m=1,2} (z_k-u_m)^{1/2} \Phi_q(\hat z_1,\hat z_k) \cr}
\eqno(b24)$$
\(b24) relates a correlation with $2N$ $\psi$'s to one
with $2(N-1)$ $\psi$'s. Thus from \(b24) we can calculate
any correlations
between $\psi$'s and $\eta$'s from those between only $\eta$'s.

We would like to point out that \(b24) is
actually over-determined. It is
crucial to check the self consistency of \(b24), namely
$\Phi_q$ calculated from different decomposition paths
(or different ways of reduction) agree with each other.
To show \(b24) is self consistent, we may
rewrite it as
$$\Phi_q(z_1..;u_1,u_2)=
\sum_{k=2} f_{u_1,u_2}(z_1,z_k) \Phi_q(\hat z_1;\hat z_k)
\eqno(b24a)$$
where
$$f_{u_1,u_2}(z,w)={1\o (z-w)^2}
{zw+u_1 u_2 -{1\o 2} (z+w)(u_1+u_2) \o
\prod_{m=1,2}
\left( (z-u_m)^{-1/2} (w-u_m)^{-1/2} \right) }
\eqno(b24b)$$
\(b24a) has a unique solution
$$\Phi_q(z_1..;u_1,u_2)=\cS
\left(f_{u_1,u_2}(z_1,z_2) f_{u_1,u_2}(z_3,z_4)...\right)
{1\o (u_1-u_2)^{2h_\eta} }
\eqno(b24c)$$
where $h_\eta$ is the dimension of $\eta$. We have explicitly
verified that the four-point correlation
$\< \psi(z_1)\psi(z_2)\eta(u_1)\eta(u_2)\>$
with $z_1\to z_2$ satisfies the conformal Ward identity if we
choose $h_\eta=1/16$. This implies that $\eta$ is a
dimension-$1/16$ primary field.
In fact $\eta$ is just the sector-changing
operator in a $c=1$ $Z_2$-orbifold model\refto{Orb,comm2}.
Such an operator is a dimension-$1/16$ primary field,
whose OPE with the $U(1)$
current (which is identified as $\psi$)
has the same structure as that in \(b21).
With this identification it is not hard to convince
oneself that $\Phi_q$ satisfies the conformal Ward identity.

Using $\eta$ we can construct the quasihole excitations in the
d-wave paired state:
$$\Phi_m(\{z_i\})
\Phi_q(\{z_i\};u_1,u_2)
\prod_k\prod_{m=1,2} (z_k-u_m)^{1/2}
\eqno(b25)$$
{}From \(b24c) we can see that \(b25) is
a single-valued and finite function of
the electron coordinates $z_i$. The wave function
describes quasiholes located at $u_m$.
The quasihole carries a fractional charge $1/2m$ in units
of electronic charge. This is because the
bound state of two such quasiholes becomes the charge-$1/m$
quasihole created by inserting a unit flux:
$$\Phi_m(z_i)
\Phi_{d}(z_i)
\prod_{k}(z_k-u)
\eqno(b26)$$
since $\lim_{u_1\to u_2} \Phi_q(z_i;u_1,u_2)
\propto \Phi_{d}(z_i)$, as one can see from \(b24c) and
\(b24b).

Notice that \(b25) can be written as a correlation between
primary fields
$$\< \prod_{m=1,2} \psi_q (u_m)
\prod_i \psi_e(z_i) e^{-i\int d^2 z \phi(z)} \>
\eqno(b27)$$
where the quasihole operator
$$\psi_q=\eta e^{i{1\o 2 \sqrt{m}}\phi}
\eqno(b28)$$
Thus as a function of $z_i$, \(b27) has the same local
structure of zeros
as the ground state wave function \(7a). The structure of zeros
is determined by the OPE \(b7) and satisfies \(7c)  if $m=3$.
Thus \(b27) is a zero-energy eigenstate of $H$ in \(7d).
We would like to stress that, when we use primary fields to
create quasiparticle excitations, the structure of zeros in the
electron wave function is not changed. This suggests
that the quasiparticle created by a primary field is a local
excitation. The above analysis also implies that the
conformal blocks generated by the quasiparticle operators
correspond to degenerate states (with zero energy).
Generally the degenerate states induce non-abelian
Berry's phases as we interchange quasiparticles,
which correspond to the non-abelian statistics
of the quasiparticles.

\head{4. The Haldane-Rezayi state}

One candidate for the $\nu=5/2$ FQH state observed in
experiments\refto{5/2} is the Haldane-Rezayi (HR)
state.\refto{HR} The HR state is a d-wave-paired
spin-singlet FQH state for spin-1/2 electrons:
$$\eqalign{
\Phi_{HR}(z_i,w_i)=&\Phi_m(z_i,w_i)\Phi_{ds}(z_i,w_i) \cr
\Phi_{ds}(z_i,w_i)=& \cA_{z,w}\left( {1\o (z_1-w_1)^2}
{1\o (z_2-w_2)^2}...\right) \cr
\Phi_m(z_i,w_i)=& \left( \prod_{i<j} (z_i-z_j)^m
\prod_{i<j} (w_i-w_j)^m \prod_{i,j} (z_i-w_j)^m \right)
e^{-{1\o 4} \sum_i (|z_i|^2+|w_i|^2)} \cr}
\eqno(11)$$
which has a filling fraction $1/m$ with $m$ an even
integer. ($m=2$ at the second Landau level gives rise to
$\nu=2+1/2=5/2$.) Here $z_i$ ($w_i$) are the coordinates
of the spin-up (-down) electrons, and the operator
$\cA_{z,w}$ performs separate antisymmetrizations
among $z_i$'s and among $w_i$'s. One can directly check
that $\Phi_{ds}$ is indeed a spin singlet, and $\Phi_m$,
when viewed as an operator, commutes with the total spin
operator.

Let us first analyze the structure of zeros in the above wave
function assuming, for simplicity, $m=2$. Let
$z_1=z_2+\de_1$ and $z_1=w_1+\de_2$, we find that in the expansion
of $\Phi_{HR}$,
$$\eqalign{
\Phi_{HR}=& \sum_{l=odd}(\de_1)^l A_{l}(z_2,..;w_1,..) \cr
\Phi_{HR}=& \sum_{n}(\de_2)^n B_{n}(z_2,..;w_1,..) \cr}
\eqno(h7b)$$
the term linear in $\de_1$ or $\de_2$ is absent:
$$A_{1}=B_{1}=0.
\eqno(h7c)$$
Therefore $\Phi$ is the exact ground state of the following
two-body Hamiltonian
$$H=V_1 \part_{z_1^*}    \de(z_1-z_2) \part_{z_1}
   +V_2 \part_{z_1^*}    \de(z_1-w_1) \part_{z_1}
\eqno(h7d)$$
since $H$ is positive definite for $V_i>0$, and
$\Phi_{HR}$ has a zero average energy.\refto{commH}
It has been checked numerically that $\Phi_{HR}$
is the unique incompressible ground state of $H$.\refto{HR}

Again we would like to know whether we can represent the pairing
wave function $\Psi_{ds}$ as a correlation of two operators
$\psi_\pm$:
$$\Phi_{ds}(z_i,w_i)=\< \prod_i\left( \psi_+(z_i)\psi_-(w_i)\right) \>
\eqno(12)$$
Since $\Phi_{ds}(z_1,w_1)\neq 0$, this motivates us to ask a simpler
question: using a CFT that contains
$$ \psi_+(z)\psi_-(w)={1\o (z_1-z_2)^{2h}}
\left(1+{2h\o c}(z_1-z_2)^2 T(z_2) + O( (z_1-z_2)^3) \right)
\eqno(13)$$
can we reproduce the wave function $\Phi_{ds}$? To answer this
question, we need to introduce the energy-momentum tensor
$$T(z)={c\o 2h}\hbox{lim}_{z_1\to z}
\left(\psi_+(z_1)\psi_-(z)- {1\o (z_1-z)^{2h}} \right) (z_1-z)^{2h-2}
\eqno(14)$$
and to check that the wave function satisfies the Ward identity \(2)
and $T$ satisfies the Virasoro algebra \(4). Note as $z_1\to w_1$,
$\Phi_{ds}$ has an expansion
$$\Phi_{ds}(z_1...;w_1...)={\Phi_{ds}(z_2...;w_2...)\o (z_1-w_1)^2}
+O(1)
$$
The absence of the $1/(z_1-w_1)$ term implies that \(14) gives us a
well-defined operator.

{}From the two point function $\Phi_{ds}(z,w)=1/(z-w)^2$,
we find the dimension of the $\psi_\pm$ is $h=1$.
By taking proper limits of the pairing wave function, we
show in appendix that $\Phi_{ds}$ indeed
satisfies the Ward identity \(2) and the Virasoro algebra
\(4), if the central charge $c$ is taken to be $c=-2$.

This result implies that $\Phi_{ds}$ can be written as a correlation
of two dimension-one operators in a $c=-2$ CFT. The $c=-2$ CFT is
closely related to a (non-unitary) minimal model, since the central
charge satisfies the condition for a minimal model\refto{BPZ}
$${\sqrt{25-c}-\sqrt{1-c}\o \sqrt{25-c}+\sqrt{1-c} }={p\o q}
\eqno(15)$$
with $p=1$ and $q=2$. Furthermore, $\psi_\pm$ (with dimension $h=1$)
are degenerate primary fields of level two\refto{BPZ}, since the
pair $(h,c)=(1,-2)$ satisfies the Kac formula \(Kac)
with $n=2$ and $m=1$. As a consequence, $\Phi_{ds}$
(as a correlation of $\psi_\pm$ in CFT) should satisfy
a second-order partial differential equation:
$$\left[ {1\o 2} \part_{z_1}^2-\sum_{i=2} ({1\o (z_1-z_i)^2}
+{1\o z_1-z_i} \part_{z_i}) - \sum_{i=1} ({1\o (z_1-w_i)^2}
+{1\o z_1-w_i} \part_{w_i}) \right] \Phi_{ds}(z_i,w_i)=0
\eqno(16)$$
and similar equations for $z_2,z_3,...$ and $w_1,w_2,...$
We have explicitly checked that $\Phi_{ds}$ indeed satisfies
the differential equation \(16).
This further confirms that $\Phi_{ds}$ can be viewed as
correlation of degenerate primary fields in a $c=-2$ CFT.

To understand the operator algebra generated from the OPE of
$\psi_\pm$, we first notice that the energy-momentum tensor $T$
is invariant under the $SU(2)$ rotation of the electron spin.
This is because $f(\xi;z_i,w_i)\equiv \<T(\xi) \prod_i
\psi_+(z_i)\psi_-(w_i) \>$ is a spin singlet when viewed
as an electron wave function (see Appendix).
Therefore our $c=-2$ CFT has  an $SU(2)$ symmetry, and
the primary fields can be arranged into $SU(2)$ multiplets,
$\psi^{s}_m$, where $s$ is the total spin and $m$ the
$S_z$ quantum number. In this notation
$\psi_\pm=\psi^{1/2}_{\pm 1/2}$. Let us
consider the leading term in the OPE of two
$\psi_+$ operators. The $SU(2)$ symmetry allows us to write
$$\psi_+(z)\psi_+(0)=C_{ {1\o 2}{1\o 2} }^1
z^{h_1-2h_{1/2}} \psi^1_1(0)+...
\eqno(17)$$
where $h_{1/2}=1$ is the dimension of $\psi_\pm$, $h_1$ and
$C_{ {1\o 2}{1\o 2} }^1$ are two constants. In CFT, OPE of
any two primary fields must be
nonzero and the leading term must be
a primary field. Therefore $\psi^1_1$ is a spin-1 primary field with
dimension $h_1$. The $SU(2)$ symmetry implies that
there exist two other primary fields $\psi^1_{0}$
and $\psi^1_{-1}$ of the same dimension.
By taking the limit $z_1\to z_2$ in the wave function
$\Phi_{sd}$, we find the leading term is proportional to $(z_1-z_2)$.
Thus $h_1-2h_{1/2}=1$ and $h_1=3$. Knowing the correlation of
$\psi_\pm$ and the OPE \(17), we can calculate any correlations
between $\psi^1_1$ and $\psi_\pm$. This, in turn, allows
us to calculate the OPE between $\psi_1^1$ and $\psi_+$:
$$\psi_+(z)\psi_1^1(0)=C_{ {1\o 2}{1} }^{3\o 2} z^{h_{3/2}-h_{1/2}-h_1}
\psi^{3/ 2}_{3/ 2}(0)+...
\eqno(18)$$
Since $\Phi_{ds}$ is antisymmetric in $z_i$, we have the expansion
$\Phi_{ds}(z_i,w_i)\propto (z_1-z_2)(z_2-z_3)(z_3-z_1)+...$ as
$z_1,z_2 \to z_3$. This implies that $\< \psi_+(z)\psi_1^1(0)...\>
\sim z^2$ as $z\to 0$ and $h_{3/2}=6$. In general the OPE
of $\psi_\pm$ generate operators with all possible values of spin.
The chiral operator algebra (the center algebra)
generated by $\psi_\pm$ contains (at least) primary fields
$\psi^s_m$, $s=$1/2, 1, 3/2,... and $m=-s$, $-s+1$,..., $s$.
The dimension $h_s$ of the spin-$s$ operator $\psi^s_m$
satisfies $h_{s+{1\o 2}}-h_s-h_{1\o 2}=2s$, which implies
$$ h_s={ (4s-1)^2-1\o 8 }
\eqno(19)$$
\(19) tells us that $\psi^s_m$ can be identified as the
level-($2s+1$) degenerate operator $\psi_{(2s+1,1)}$ in
the $c=-2$ model. (Here we use the notation in \Ref{BPZ}.)
In contrast to other minimal models, our $c=-2$ CFT contains
infinite number of primary fields and has a
global $SU(2)$ symmetry.

The operator algebra of $\psi^s_m$ fields determines the correlations
between those operators. However to calculate the correlations
between $\psi_\pm$ we do not need to know the full structure
of the operator algebra. In fact the following operator
algebra
$$\eqalign{
\part_{\bar z}\psi_\pm=& 0\cr
\{ \psi_a(z_1),\psi_a(z_2)\}=& 0, \ \ \ \ a=+,- \cr
\psi_+(z)\psi_-(0)=& {1\o z^2}+O(1) \cr}
\eqno(20)$$
completely determines the correlations between $\psi_\pm$,
since the algebra \(20) completely determines the poles and their
residues in the correlation.

Now let us introduce the electron operator
$$\psi_{e\pm}(z)=\psi_\pm (z) e^{i\sqrt{m} \phi(z)}.
\eqno(h5)$$
Then the wave function $\Phi_{HR}$ in \(11) can be
written as
$$\Phi_{HR}=\<\prod_i \left[\psi_{e+}(z_i)  \psi_{e-}(w_i) \right]
e^{-i{1\o \sqrt{m}} \int d^2 z \phi} \>,
\eqno(h6)$$
and the constraint \(h7c) on zeros of $\Phi_{HR}$
can be derived
from the OPE of $\psi_{e\pm}$. Let $z_1=z_2+\de_1$ and
$z_1=w_1+\de_2$ and assume $m=2$; we find
$$\eqalign{
 \psi_{e+}(z_1) \psi_{e+}(z_2) =&
 (\de_1)^3 \psi_1^1(z_2) e^{2i\sqrt{m} \phi(z_2)} \cr
 \psi_{e+}(z_1) \psi_{e-}(w_1) =&
  e^{2i\sqrt{m} \phi(w_1)} + O((\de_2)^2) \cr}
\eqno(h7)$$
The absence of terms linear in $\de_1$ and $\de_2$
implies \(h7c).

To study the quasiparticles in the HR state,
we need to extend the operator algebra
\(20) to include the disorder operator $\eta$.
The $c=-2$ CFT contains another level-2 degenerate primary field
$\psi_{(1,2)}$ which has a dimension $-1/8$. It is natural to
identify this operator as the disorder operator $\eta$. From
the fusion rule in the $c=-2$ model
$\psi_{(2,1)}(z)\psi_{(1,2)}(0)\sim z^{-1/2}\psi_{(2,2)}(0)$, we find
$$\psi_\pm(z)\eta(0)\sim z^{-1/2} (\t\eta_\pm(0)+O(z))
\eqno(21)$$
Note that as we move $\psi_\pm$ around $\eta$,
the correlation obtains a
minus sign which commutes with the $SU(2)$ spin rotation. Thus
we expect $\eta$ to be a spin-singlet operator.

To calculate the correlations between
$\psi_\pm$ and two $\eta$'s \refto{com} we note that again
$$\prod_{m=1,2} (z_1-u_m)^{1/2}
\<\prod_{i=1}\left(\psi_+(z_i) \psi_-(w_i)\right)\prod_{m=1,2} \eta(u_m)\>
\eqno(22)$$
is a holomorphic function of $z_1$ which has only poles at $z_1=w_i$.
We also note that, as a function of $z_1$,
$$\Phi_q(z_1..;w_1..;u_1,u_2)\equiv
\<\prod_{i=1}\left(\psi_+(z_i) \psi_-(w_i)\right)\prod_{m=1,2} \eta(u_m)\>
\eqno(23)$$
has a pure second-order pole (no first order pole)
at $z_1=w_k$ with a residue
$(-)^{k+1}\Phi_q(\hat z_1;\hat w_k)$, where $\Phi_q(\hat z_1;\hat w_k)$
is $\Phi_q(z_1..;w_1..;u_1,u_2)$ with variables $z_1$ and $w_k$ removed.
This implies that the second-order pole at $z_1=w_k$ in \(22)
has a residue $(-)^{k+1}\Phi_q(\hat z_1;\hat w_k)\prod_{m=1,2}
(w_k-u_m)^{1/2}$ and the first-order pole has a residue
$\part_{z_1}(-)^{k+1}\Phi_q(\hat z_1;\hat w_k)\prod_{m=1,2} (z_1-u_m)^{1/2}$
evaluated at $z_1=w_k$.
Thus we obtain the following relation between correlations $\Phi_q$:
$$\eqalign{
 &\Phi_q(z_1..;w_1..;u_1,u_2) \cr
=&\prod_{m=1,2} (z_1-u_m)^{-1/2}\sum_k (-)^{k+1}
\left({1\o (z_1-w_k)^2}+{1\o 2(z_1-w_k)}\sum_{m=1,2} {1\o w_k-u_m}\right) \cr
 &\times \prod_{m=1,2} (w_k-u_m)^{1/2} \Phi_q(\hat z_1;\hat w_k) \cr}
\eqno(24)$$
It relates a correlation with $N$ pairs of $\psi_\pm$ to the one
with $N-1$ pairs, and determines any correlations
between $\psi_\pm$'s and two $\eta$'s from those between
only $\eta$'s.

Again \(24) is over determined and it is
crucial to check the self-consistency of \(24).
In fact \(24) can be rewritten as
$$\Phi_q(z_1..;w_1..;u_1,u_2)=
\sum_k (-)^{k+1}f_{u_1,u_2}(z_1,w_k) \Phi_q(\hat z_1;\hat w_k)
\eqno(24a)$$
where $f_{u_1,u_2}(z,w)$ is given in \(b24b).
\(24a) has a unique solution
$$\Phi_q(z_1..;w_1..;u_1,u_2)=\cA_{z,w}
\left(f_{u_1,u_2}(z_1,w_1) f_{u_1,u_2}(z_2,w_2)...\right)
{1\o (u_1-u_2)^{2h_\eta} }
\eqno(24c)$$
Since $f_{u_1,u_2}(z,w)=f_{u_1,u_2}(w,z)$,
we see that $\Phi_q$ is a spin singlet, which
implies that $\eta$ is a spin-singlet operator,
as we have expected. In the Appendix we will show that
$\Phi_q$ satisfies the conformal Ward identity if we
choose $h_\eta=-1/8$. Thus $\eta$ is a dimension-($-1/8$)
primary field, which agrees with general considerations
for the $c=-2$ CFT.

Using $\eta$, we can write down
the wave function of the quasihole
excitations in the HR state:
$$\Phi_m(\{z_i;w_j\})
\Phi_q(\{z_i;w_j\};u_1,u_2)
\prod_k\prod_{m=1,2} (z_k-u_m)^{1/2}(w_k-u_m)^{1/2}
\eqno(25)$$
{}From \(24c) we can see that \(25) is single-valued
and non-singular as a function of
the electron coordinates $z_i$ and $w_j$.
The wave function describes quasiholes whose locations are
described by
the parameters $u_m$. The quasihole carries $1/2m$ unit of
electron charge, which is induced by the factor
$\prod_{k,m}\left( (z_k-u_m)^{1/2}(w_k-u_m)^{1/2}\right)$. The
bound state of two such quasiholes becomes the charge-$1/m$
quasihole:
$$\Phi_m(z_i;w_j)
\Phi_{ds}(z_i;w_j)
\prod_{k}(z_k-u_m)(w_k-u)
\eqno(26)$$
since $\lim_{u_1\to u_2} \Phi_q(z_i;w_j;u_1,u_2)
\propto \Phi_{ds}(z_i;w_j)$ (see \(24c).
Again since the quasiholes are created by insertion
of primary fields, the quasihole wave function \(25)
is a zero-energy eigenstate of $H$ in \(h7d).
The system also contains neutral spin-1/2 excitations
which can be viewed as a bound state of an electron
and $m$ charge-$1/m$ quasiholes. Other excitations can be viewed
as bound states of charge $1/2m$ spin-singlet excitations and
neutral spin-1/2 excitations.\refto{comm3}

\head{5. D-wave-paired spin-triplet FQH state}

The d-wave-paired spin-triplet FQH state for spin-1/2 electrons is
given by:
$$\eqalign{
\Phi_{DL}(z_i,w_i)=&\Phi_{lmn}(z_i,w_i)\Phi_{dt}(z_i,w_i) \cr
\Phi_{dt}(z_i,w_i)=& \cS_{z,w}\left( {1\o (z_1-w_1)^2}
{1\o (z_2-w_2)^2}...\right) \cr
\Phi_{lmn}(z_i,w_i)=& \left( \prod_{i<j} (z_i-z_j)^l
\prod_{i<j} (w_i-w_j)^m \prod_{i,j} (z_i-w_j)^n \right)
e^{-{1\o 4} \sum_i (|z_i|^2+|w_i|^2)} \cr}
\eqno(11t)$$
which has a filling fraction ${l+m-2n\o lm-n^2}$
with $l,m$ odd integers. Here
$z_i$ ($w_i$) are the coordinates of the spin-up
(-down) electrons, and the operator $\cS_{z,w}$
performs separate symmetrizations between $z_i$'s
and between $w_i$'s. $\Phi_{DL}$ is not a spin singlet
and there is no spin rotation symmetry.
It is more natural to view $\Phi_{DL}$ as a wave function
for double layer FQH systems which involve
interlayer pairing. Now $z_i$ and $w_i$ are
electron coordinates in the two layers.
We will call the FQH state in \(11t) the DL state.

Again we would like to know whether we can represent the pairing
wave function $\Psi_{dt}$ as a correlation of two primary fields
$\psi_\pm$:
$$\Phi_{dt}(z_i,w_i)=\< \prod_i\left( \psi_+(z_i)\psi_-(w_i)\right) \>
\eqno(12t)$$
Note $\Phi_{dt}$ has the following expansion as $z_1\to w_1$
$$\Phi_{dt}(z_1...;w_1...)={\Phi_{dt}(z_2...;w_2...)\o (z_1-w_1)^2}
+O(1)
$$
This motivates us to assume
$$ \psi_+(z)\psi_-(w)={1\o (z_1-z_2)^{2h}}
\left(1+{2h\o c}(z_1-z_2)^2 T(z_2) + O( (z_1-z_2)^3) \right)
\eqno(13t)$$
with the dimension of $\psi_\pm$ being $h=1$. Thus we may
introduce the energy-momentum tensor through
$$T(z)={c\o 2h}\hbox{lim}_{z_1\to z}
\left(\psi_+(z_1)\psi_-(z)- {1\o (z_1-z)^{2h}} \right) (z_1-z)^{2h-2}
\eqno(14t)$$

We have directly checked that the wave function $\Phi_{dt}$
satisfies the Ward identity \(2)
and $T$ satisfies the Virasoro algebra \(4) if we choose $c=2$.
Therefore the
 $\Phi_{dt}$ can be written as a correlation of
a dimension-1 operator in a $c=2$ conformal theory.
In fact, one may consider a
$U(1)\times U(1)$ Gaussian model which has $c=2$. Let
$$j_\pm={1\o \sqrt{2}}(j_1\pm ij_2)
\eqno(15t)$$
where $j_1$ and $j_2$ are the currents of the two $U(1)$ models.
One can easily check that the correlations of $j_\pm$ reproduce
$\Phi_{dt}$. Thus we can identify $\psi_\pm$ as $j_\pm$.

The wave function $\Phi_{dt}$ can be derived from
the following operator algebra
$$\eqalign{
\part_{\bar z}\psi_\pm=& 0\cr
[ \psi_a(z_1),\psi_a(z_2)]=& 0, \ \ \ \ a=+,- \cr
\psi_+(z)\psi_-(0)=& {1\o z^2}+O(1) \cr}
\eqno(t20)$$
To study the quasiparticles in the DL state, we need to extend the
operator algebra \(t20) to include the disorder operator $\eta$.
Again let us try
$$\psi_\pm(z)\eta(0)\sim z^{-1/2} (\t\eta_\pm(0)+O(z))
\eqno(t21)$$

The correlations $\Phi_q$ between $\psi_\pm$ and two
$\eta$'s \refto{com} can be calculated from \(t20) and \(t21).
We find that $\Phi_q$ satisfies
$$\eqalign{
 &\Phi_q(z_1..;w_1..;u_1,u_2) \cr
=&\prod_{m=1,2} (z_1-u_m)^{-1/2}\sum_k
\left({1\o (z_1-w_k)^2}+{1\o 2(z_1-w_k)}\sum_{m=1,2}
{1\o w_k-u_m}\right) \cr
 &\,\,\,\,\, \times \prod_{m=1,2} (w_k-u_m)^{1/2}
\Phi_q(\hat z_1;\hat w_k) \cr}
\eqno(t24)$$
Again \(t24) can be rewritten as
$$\eqalign{
&\Phi_q(z_1..;w_1..;u_1,u_2)\cr
=& \sum_k f_{u_1,u_2}(z_l,w_k) \Phi_q(\hat z_l;\hat w_k) \cr
=&\cS_{z,w}
\left(f_{u_1,u_2}(z_1,w_1) f_{u_1,u_2}(z_2,w_2)...\right)
{1\o (u_1-u_2)^{2h_\eta} } \cr}
\eqno(t24a)$$
where $f_{u_1,u_2}(z,w)$ is given in \(b24b).
The fact that \(t24a) has an unique solution
implies that the OPE \(t20) and \(t21) are self consistent.
One can show that $\Phi_q$ satisfies the conformal
Ward identity if we choose $h_\eta=1/8$. This implies that
$\eta$ is a dimension-$1/8$ primary field. Again
$\eta$ can be identified as the sector-changing operator
in the $c=2$ orbifold model.

The quasihole excitations in the DL
state is given by
$$\Phi_m(\{z_i;w_j\})
\Phi_q(\{z_i;w_j\};u_1,u_2)
\prod_k\prod_{m=1,2} (z_k-u_m)^{1/2}(w_k-u_m)^{1/2}
\eqno(t25)$$
\(t25) is a single-valued and finite function of
the electron coordinates $z_i$ and $w_j$. The wave function
describes quasiholes located at $u_m$ which
carry electric charge $1/2m$. The
bound state of two such quasiholes becomes the
charge-$1/m$
quasihole created by inserting a unit flux:
$$\Phi_m(z_i;w_j)
\Phi_{dt}(z_i;w_j)
\prod_{k}(z_k-u)(w_k-u)
\eqno(t26)$$
This is because $\lim_{u_1\to u_2} \Phi_q(z_i;w_j;u_1,u_2)
\propto \Phi_{dt}(z_i;w_j)$ (see \(t24a)).

\head{6. Conclusions and Discussions}

In this paper we have proposed a new approach to explore
the relationship between FQHE and CFT. Instead of
generating FQH wave functions from correlations in a
CFT chosen {\it a priori}, we start with a certain FQH
many-body wave function and ask whether it can be
consistently viewed as a correlation in appropriate
CFT. Following techniques often used in CFT, we find
that by choosing some simple ansatz for the OPE of the
electron operator, whose consistency can be checked
later, conformal Ward identities can be turned into partial
differential equations relating the $N$-body to less-body
wave functions, if the given FQH wave function is to
be identified with correlations in a (chiral) CFT.
The conformal dimension, $h$, of the primary field
that is to be identified with
(the non-Laughlin part of) the electron
operator and the central charge, $c$,
of the CFT are determined in the course
of verifying the conformal Ward identities.
If the pair $(h,c)$ happen to satisfy the Kac
formula, the primary field must be a degenerate
one and their correlations, according to
general principles of CFT, must satisfy certain
partial differential equations called the null
vector condition. Whether the
given FQH wave function satisfies these
equations is an independent check
for their identification with correlation
functions in the CFT. We have explicitly
shown that several many-body FQH wave functions,
such as p-wave and d-wave paired states
for spinless electrons and d-wave paired
spin singlet and triplet for spin-1/2 electrons,
indeed pass the check for conformal Ward
identities and for the null vector condition.
We feel that passing through these nontrivial
checkings should not be a mere accident; it
really implies the existence of a conformal
operator product algebra (the center algebra),
which is generated by the OPE of the electron
operator(s) as primary field, hidden in {\it these}
FQH wave functions. We propose to use this
conformal operator algebra to characterize the topological
order in the corresponding FQH states. Namely the
quantum numbers of the quasiparticle excitations
in these FQH states, the ground state degeneracy
on a torus as well as the spectrum of edge states
should be determined from the knowledge of the
associated hidden conformal operator product
algebra.

Leaving the discussions on other topological
properties to further research, in this
paper we have tried only to determine
the quantum numbers of quasiparticle
excitations. To study quasiparticle
excitations we need to consistently extend
the operator product algebra that we obtained
from the electron operator(s) to include
disorder operators. This is a highly
nontrivial task in a {\it chiral}
CFT. We have not been able to give
a systematic procedure or a criterion
for the completeness for such extension.
By guesswork and some consistency check,
we have been able to include certain
simple disorder operators and use them to
make predictions on quantum numbers
of part of quasiparticles and calculate their
wave functions. It seems that
a more systematic study needs
considerations of how to incorporate
more irreducible representations of
the operator algebra (the center algebra)
generated by the electron operator(s)
to form a (minimal) closed fusion-rule algebra.
Or one may try to find a bigger underlying
operator algebra that organize all
states of the FQH system, the ground
state and all its excitations, into
a single irreducible representation.

Finally we feel it is necessary to make the
following cautious remark. It is not clear to us
whether every conceivable
FQH wave function must be compatible with
conformal invariance or conformal algebra.
For example, our construction does not apply
to f-wave paired state for spinless electrons,
though the f-wave paired state is very similar
to the p-wave paired state.
So we feel the operator algebra generated
from the electron operator(s) and its
extension by including disorder operators
perhaps are more fundamental. Whether the algebra
generated by electron (the center algebra)
forms a representation of Virasoro
algebra or not is not very important.
In fact, in above sections we have
seen in many cases these operator algebras
alone are sufficient to determine relevant
wave functions as holomorphic correlations.
The existence of the Virasoro algebra simply
organizes the operators in the center algebra
into Verma modules and make it easier to study
the structure of the center algebra.

We would like to thank Aspen Center of Physics
for hospitality, where this work was initiated.
XGW is supported by NSF grant No. DMR-91-14553
and YSW by NSF grant No. PHY-9309458.

\head{ Appendix }
\def\G{\Phi_{ds}}

Here we like to show that, for the d-wave spin-singlet
wave function $\G$, the energy-momentum tensor $T$
defined in \(14) satisfies the Ward identity \(2) and
the Virasoro algebra \(4)
if $c=-2$. The dimension of $\psi_\pm$ is taken to be $h=1$.

Two relations of $\G$ are very useful:
$$\G=\sum_{k} (-)^{i+k}\G(z_i;w_k)
\G(\hat z_i ;\hat w_k)
\eqno(A1)$$
$$\G=\sum_{k'<k} (-)^{k+k'+1}\G(z_1,z_2;w_{k'},w_k)
\G(\hat z_1,\hat z_2 ;\hat w_{k'},\hat w_k)
\eqno(A2)$$
where $\G(\hat z_i ;\hat w_k)$ is $\G(z_1,..;w_1,..)$ with $z_i$ and
$w_k$ variables removed, and $\G$ denotes $\G(z_1,..;w_1,..)$.

To check the Ward identity, we first calculate
$\< T(\xi) \prod_i [\psi_+(z_i)\psi_-(w_i)]\>$. Consider
$$\G(\xi+\de,z_1,..;\xi,w_1,..)=\de^{-2}\G+\sum_k
{(-)^k\o (\xi+\de-w_k)^2} \G(z_1..;\xi,\hat w_k)
\eqno(A3)$$
Taking the limit $\de \to 0$ and subtracting out the $\de^{-2}$ term,
we find
$$\eqalign{
 & \< T(\xi) \prod_i [\psi_+(z_i)\psi_-(w_i)]\> \cr
=&{c\o 2}\sum_k {(-)^k\o (\xi-w_k)^2} \G(z_1..;\xi,\hat w_k) \cr
=& -{c\o 2}
\sum_{i,k} (-)^{i+k}\G(\xi,z_i)\G(\xi;w_k)\G(\hat z_i;\hat w_k)\cr}
\eqno(A4)$$
This expression tells us that when viewed as a function of $z_i$
and $w_j$,
$$\< T(\xi) \prod_i [\psi_+(z_i)\psi_-(w_i)]\>
=-{c\o 2}\cA_{z,w}\left( {1\o (z_1-\xi)^2(w_1-\xi)^2} {1\o
(z_2-w_2)^2}...\right)
\eqno(A4a)$$
is a spin-singlet function and $T(\xi)$ is a spin-singlet operator.

{}From the relation
$$\eqalign{
 &\sum_i \left({1\o (\xi-z_i)^2} +{1\o \xi-z_i}\part_{z_i}
	      +{1\o (\xi-w_i)^2} +{1\o \xi-w_i}\part_{w_i} \right)\G
\cr
=&\sum_{i,k}\left({1\o (\xi-z_i)^2} +{1\o \xi-z_i}\part_{z_i}\right)
(-)^{i+k}\G(z_i;w_k)\G(\hat z_i;\hat w_k) \cr
 &+\sum_{i,k}\left({1\o (\xi-w_k)^2} +{1\o \xi-w_k}\part_{w_k}\right)
(-)^{i+k}\G(z_i;w_k)\G(\hat z_i;\hat w_k) \cr }
\eqno(A5)$$
one can directly check, by comparing with \(A4), that
the Ward identity \(2) is indeed satisfied
by the $\G$ wave function if $c=-2$.

To check the Virasoro algebra \(4) we need to calculate
$\< T(\xi) T(\t\xi) \prod_i [\psi_+(z_i)\psi_-(w_i)]\>$.
Using \(A2) then \(A1) we can rewrite
$$\eqalign{
 &\G(\xi_1,\t\xi_1,z_1..;\xi,\t\xi,w_1..)  \cr
=&\G(\xi_1,\t\xi_1;\xi,\t\xi)\G
 +\sum_{i,k}\G(\xi_1,\t\xi_1;\xi,w_k)\G(z_i;\t\xi)\G(\hat z_i;\hat w_k)
   (-)^{i+k+1} \cr
&+\sum_{i,k}\G(\xi_1,\t\xi_1;\t\xi,w_k)\G(z_i;\xi)\G(\hat z_i;\hat w_k)
   (-)^{i+k} \cr
&+\sum_{i'<i} \sum_{k'<k}\G(\xi_1,\t\xi_1;w_{k'},w_k)
\G(z_{i'},z_i;\xi,\t\xi)\G(\hat z_{i'},\hat z_i;\hat w_{k'},\hat w_k)
   (-)^{i+i'+k+k'} \cr }
\eqno(A6)$$
Let $\xi_1=\xi+\de_1$ and $\t\xi_1=\t\xi+\de_2$, we find, as
$\de_{1,2}\to 0$,
$$\G(\xi_1,\t\xi_1,z_1..;\xi,\t\xi,w_1..)={A\o \de_1^2\de_2^2}
+{B\o \de_1^2}+{C\o\de_2^2}+O(1)
$$
The singular terms are cancelled by the subtraction in the definition
of $T$ (see \(14)). The finite term gives
$$\eqalign{
& {4\o c^2}\< T(\xi) T(\t\xi) \prod_i [\psi_+(z_i)\psi_-(w_i)]\>\equiv
{4\o c^2} F(\xi,\t\xi;z_1..;w_1..)  \cr
=&-{\G\o (\xi-\t\xi)^4}
+\sum_{i,k}{(-)^{i+k} \o (\xi-\t\xi)^2}
\left( \G(\t\xi;w_k)\G(z_i;\xi)+\G(\xi;w_k)\G(z_i;\t\xi) \right)
\G(\hat z_i;\hat w_k)  \cr
&+\sum_{i'<i,k'<k}\G(\xi,\t\xi;w_{k'},w_k)
\G(z_{i'},z_i;\xi,\t\xi)\G(\hat z_{i'},\hat z_i;\hat w_{k'},\hat w_k)
   (-)^{i+i'+k+k'} \cr }
\eqno(A7)$$
The above is to be compared with the right hand side of \(4):
$$\eqalign{
       &F'(\xi,\t\xi;z_1..;w_1..) \cr
\equiv &\sum_k
\left( {1\o (\xi -z_k)^2} +{1\o \xi -z_k} \part_{z_k}
      +{1\o (\xi -w_k)^2} +{1\o \xi -w_k} \part_{w_k}
\right)
\< T(\t \xi) \prod_i\psi_+(z_i) \psi_-(w_i) \>\cr
&+\left( {2\o (\xi -\t \xi)^2}+{1\o \xi -\t \xi} \part_{\t \xi} \right)
 \< T(\t \xi) \prod_i \psi_+(z_i) \psi_-(w_i) \>
 +{c/2 \o (\xi-\t \xi)^4} \G \cr}
\eqno(A8)$$
Let us compare the structure of poles in $F$ and $F'$. With the help
of \(A1) and \(A2), we find, in the limit $\t\xi\to\xi$,
$$F=-{1\o (\xi-\t \xi)^4} \G
 +\left( {2\o (\t\xi-\xi)^2}+{1\o \t\xi-\xi} \part_{\xi} \right)
\sum_{i,k} (-)^{i+k}\G(\xi;z_i)\G(\xi;w_k)\G(\hat z_i;\hat w_k)
 +O(1)
\eqno(A9)$$
and, in the limit $\xi\to z_m$,
$$F=\left( {1\o (\xi -z_m)^2} +{1\o \xi -z_m} \part_{z_m}\right)
\sum_{i,k} (-)^{i+k}\G(\t\xi;z_i)\G(\t\xi;w_k)\G(\hat z_i;\hat w_k)+O(1)
\eqno(A10)$$
and similar result for $\xi\to w_m$.

Using \(A4) one can show that
$F-F'$, as a function of $\xi$, has no singularities if we choose
$c=-2$. Thus $F-F'$ is a
polynomial of $\xi$ since $F$ and $F'$ are holomorphic functions of
$\xi$. Knowing $F,F'\to 0$ as $\xi\to \infty$, we find $F=F'$ and the
$T$ in \(14) satisfies the Virasoro algebra of $c=-2$
for the wave function $\G$.

To show $\Phi_q$ in \(24c) satisfies the conformal Ward identity
we need to use the following property of $\Phi_q$
$$\Phi_q
=\sum_{k} (-)^{i+k}f_{u_1,u_2}(z_i,w_k)
\Phi_q(\hat z_i ;\hat w_k)
\eqno(A11)$$
Applying \(A11) twice we find
$$\eqalign{
 &\Phi_q(\xi+\de,z_1,..;\xi,w_1,..;u_1,u_2) \cr
=&f_{u_1,u_2}(\xi+\de,\xi)\Phi_q+
\sum_{i,k} (-)^{1+i+k}f_{u_1,u_2}(\xi+\de,w_k)f_{u_1,u_2}(\xi,z_i)
\Phi_q(\hat z_i ;\hat w_k)  \cr}
\eqno(A12)$$
Taking $\de\to 0$ and subtracting out
the $\de^{-2}$ term we obtain
$$\eqalign{
&\< T(\t \xi) \eta(u_1)\eta(u_2)
\prod_i\psi_+(z_i) \psi_-(w_i) \>  \cr
=&{c\o 2} {1\o 8}{(u_1-u_2)^2 \Phi_q\o (u_1-\xi)^2(u_2-\xi)^2}
-{c\o 2}\sum_{i,k} (-)^{i+k}
 f_{u_1,u_2}(\xi,w_k)f_{u_1,u_2}(\xi,z_i)
\Phi_q(\hat z_i ;\hat w_k)  \cr}
\eqno(A13)$$
The right hand side of \(A13) can be shown to be equal to
$$\eqalign{
&\sum_i \left({1\o (\xi-z_i)^2} +{1\o \xi-z_i}\part_{z_i}
+{1\o (\xi-w_i)^2} +{1\o \xi-w_i}\part_{w_i} \right)\Phi_q  \cr
&+\left({h_\eta\o (\xi-u_1)^2} +{1\o \xi-u_1}\part_{u_1}
+{h_\eta\o (\xi-u_2)^2} +{1\o \xi-u_2}\part_{u_2} \right)\Phi_q  \cr}
\eqno(A14)$$
if we choose $c=-2$ and $h_\eta=-1/8$. Thus $\eta$ is a
primary field with dimension $-1/8$.
In fact \(A14) can be written as
$$
f_{du}\Psi_q+\sum_{i,j} (-)^{i+j} f_{dij} \Phi_q(\hat z_i ;\hat w_j)
\eqno(A15)$$
where
$$f_{du}=(u_1-u_2)^{2h_\eta} \left(
{h_\eta\o (\xi-u_1)^2} +{1\o \xi-u_1}\part_{u_1}
+{h_\eta\o (\xi-u_2)^2} +{1\o \xi-u_2}\part_{u_2} \right) (u_1-u_2)^{-2h_\eta}
\eqno(A16)$$
and
$$\eqalign{
f_{dij}=&\left(
{1\o (\xi-z_i)^2} +{1\o \xi-z_i}\part_{z_i}
+{1\o (\xi-w_j)^2} +{1\o \xi-w_j}\part_{w_j}
+{1\o \xi-u_1}\part_{u_1} +{1\o \xi-u_2}\part_{u_2}
\right) f_{u_1,u_2}(z_i,w_j) \cr}
\eqno(A17)$$
For $h_\eta=-1/8$ one can show that
$$f_{du}=-{1\o 8}{(u_1-u_2)^2\o (u_1-\xi)^2(u_2-\xi)^2}
\eqno(A18)$$
and
$$f_{dij}=f_{u_1,u_2}(\xi,w_j)f_{u_1,u_2}(\xi,z_i)
\eqno(A19)$$
which leads to the equality between \(A13) and \(A14).

\references

\refis{h}
F.D.M. Haldane, \prl 51, 605, 1983;
B. I. Halperin, \prl 52, 1583, 1984.
S. Girvin, \prb 29, 6012, 1984; A.H. MacDonald and D.B. Murray,
 \prb 32, 2707, 1985; M.P.H. Fisher and D.H. Lee, \prl 63, 903,
1989; J.K. Jain, \prl 63, 199, 1989; \pr B41, 7653, 1991.

\refis{mat}
B. Blok and X.G. Wen, \pr B42, 8133, 1990; \pr B42, 8145, 1990;
N. Read, \prl 65, 1502, 1990;
J. Fr\"ohlich and A. Zee, \np  B364, 517, 1991;
Z.F. Ezawa and A. Iwazaki, \pr B43, 2637, 1991;
X.G. Wen and A. Zee, \prb 46, 2290, 1992.

\refis{eff}
S. M. Girvin and A. H. MacDonald, \prl 58, 1252, 1987;
S. C. Zhang, T. H. Hansson and S. Kivelson, \prl 62, 82, 1989;
N. Read, \prl 62, 86, 1989; Z.F. Ezawa and A. Iwazaki, \pr B43, 2637, 1991.

\refis{rd}
X.G. Wen,
\journal Int. J. Mod. Phys. B, 2, 239, 1990;
\pr B40, 7387, 1989;
 X.G. Wen and Q. Niu,  \pr B41, 9377, 1990.

\refis{W}
X.G. Wen,  \prl 66, 802, 1991; B. Blok and X.G. Wen,
\np B374, 615, 1992.

\refis{witten}
E. Witten, \journal Comm. Math. Phys., 121, 351, 1989;
J. Fr\"ohlich and C. King, \journal Comm. Math. Phys., 126, 167, 1989;

\refis{MR}
G. Moore and N. Read, \np B360, 362, 1991.

\refis{Fubini}
S. Fubini, \journal Int. J. Mod. Phys., A5, 3553, 1990;
S. Fubini and C.A. L\"utken, \journal Mod. Phys. Lett., A6, 487, 1991.

\refis{Naples}
G. Christofano, G. Maiella, R. Musto and F. Nicodemi,
\pl B262, 88, 1991;
\journal Mod. Phy. Lett., A6, 1779, 1991;
{\it ibid} {\bf A6}, 2985 (1991); {\it ibid} {\bf A7}, 2583 (1992).

\refis{HR1}
F.D.M. Haldane and E.H. Rezayi, \prb 31, 2529, 1985.

\refis{cg}
D.P. Li, \journal Int. J. Mod. Phys., B7, 2655, 1993;
{\it ibid} {\bf B7}, 2779, 1993.
E. Keski-Vakkuri, and X.G. wen, preprint MIT-CTP-2197 (hep-th/9303155).

\refis{edge}
X.G. Wen, \journal Int. J. Mod. Phys., B6, 1711, 1992;
J. Fr\"ohlich and T. Kerler, \np B354, 369, 1991.

\refis{Wp} X.G. Wen, \prl 70, 355, 1993.

\refis{HR} F.D.M. Haldane and E.H. Rezayi, \prl 60, 956, 1988;
{\bf 60}, E1886 (1988).

\refis{BPZ}
A.A. Belavin, A.M. Polyakov and A.B. Zamolochikov,
\np B241, 333, 1984.

\refis{book}
``{\it Conformal Invariance and Applications to Statistical
Mechanics}'', ed. C. Itzykson, H. Saleur and J.B. Zuber,
World Scientific, 1988.

\refis{RR} N. Read and E.H. Rezayi, Yale and CSU preprint, May 1993.

\refis{commH} It is not hard to verify that the inclusion of
the U(1) part, $\Phi_m$, does not change this statement, i.e.
the total wave function is still a zero-energy eigenstate
of the Hamiltonian.

\refis{com} When there are more then two $\eta$ operators, the correlation
may have several conformal blocks. We limit ourselves to consider
only two  $\eta$ operators, so that only one conformal block is
involved in the correlation.

\refis{comm3}
The HR state was first considered from the CFT point of view in
\ref{MR}, where a CFT of ghosts plus two  Gaussian models
are invoked. In our
treatment, we have an explicit global $SU(2)$ symmetry;
and we have discussed the quantum numbers of the quasiparticles.

\refis{Orb}
L. Dixon, J.A. Harvey, C. Vafa, and E. Witten,
\np B261, 620, 1985;
\np B274, 285, 1986.
L. Dixon, D. Friedan, E. Martinec ans S. Shenker,
\np B282, 13, 1987.

\refis{comm2} Our $c=1$ $Z_2$-orbifold model here
is the one with radius $R=\infty$. Actually it
can be viewed as a $R^1/Z_2$ current algebra model.

\refis{5/2}
R.L. Willet, J.P. Eisenstein, H.L. St\"ormer, D.C. Tsui,
A.C. Gossard and J.H. English, \prl 59, 1776, 1987.

\refis{GWW}
M. Greiter, X.G. Wen amd F. Wilczek, \prl 66, 3605, 1991;
\np B374, 567, 1992.

\endreferences

\end